\documentstyle[12pt,graphicx]{article}
\newcommand{\be}{\begin{equation}}
\newcommand{\ee}{\end{equation}}
\newcommand{\bn}{\begin{eqnarray}}
\newcommand{\en}{\end{eqnarray}}
\def\al{\alpha}
\def\th{\theta}
\def\Th{\Theta}
\def\be{\beta}
\def\Ga{\Gamma}

\def\lam{\lambda}
\def\der{\partial}

\begin{document}
\title{Exact Solutions of the Klein-Gordon Equation in the Presence of a Dyon, Magnetic Flux and Scalar Potential in the Specetime of Gravitational Defects}
\author{A. L. Cavalcanti de Oliveira \thanks{E-mail: alo@fisica.ufpb.br} \ 
and E. R. Bezerra de Mello \thanks{E-mail: emello@fisica.ufpb.br}
\\
Dept. de F\'{\i}sica-CCEN. Universidade Federal da Para\'{\i}ba\\
58.059-970, J. Pessoa, PB. C. Postal 5.008. Brazil}
\maketitle
\begin{abstract}
In this paper we analyse the relativistic quantum motion of a charged spin$-0$ particle in the presence of a dyon, Aharonov-Bohm magnetic field and scalar potential, in the spacetimes produced by an idealized cosmic string and global monopole. In order to develop this analysis, we assume that the dyon and the Aharonov-Bohm magnetic field are superposed to both gravitational defects. Two distinct configurations for the scalar potential, $S(r)$, are considered: $i)$ the potential proportional to the inverse of the radial distance, i.e.,  $S\propto1/r$, and $ii)$ the potential proportional to this distance, i.e., $S\propto r$. For both cases the center of the potentials coincide with the dyon's position. In the case of the cosmic string the Aharonov-Bohm magnetic field is considered along the defect, and for the global monopole this magnetic field pierces the defect. The energy spectra are computed for both cases and explicitly shown their dependence on the electrostatic and scalar coupling constants. Also we analyse scattering states of the Klein-Gordon equations, and show how the phase shifts depend on the geometry of the spacetime and on the coupling constants parameter.
\end{abstract}

\newpage
\renewcommand{\thesection}{\arabic{section}.}
\section{Introduction}
The influence of the gravitational field on quantum mechanical systems has attracted attention in particle physics several years ago. In this way, the analysis of the hydrogen atom in curved spacetime has been considered in \cite{AS,Parker,Parker1}. In \cite{Parker} was shown that the shifts in the energy spectrum caused by local curvature is different from the usual gravitational Doppler shift. On the other hand, this shift is appreciable only in the region of strong gravitational field. Recently the analysis of the influence of the topology of the spacetime on the energy spectrum of the hydrogen atom, has been considered in a nonrelativistic \cite{GB} and a relativistic \cite{GB1} point o views. In these papers the hydrogen atom is placed in the spacetime produced by an idealized linear cosmic string, and a point-like global monopole. Different from the previous analysis, in these recent investigations the energy spectra associated with the hydrogen atom could be exactly calculated. 

The fermion-dyon system has been analyzed in \cite{Zhang,Zhang1} under a relativistic point of view in a flat spacetime. In these papers the energy spectrum associated with this system was shown to be similar to the hydrogen atom one if the product between the electric charge of the particle, $e$, with the magnetic charge of the monopole, $g$, is an integer number. The nonrelativistic quantum analysis of charged particle in the presence of a magnetic monopole and in the global monopole spacetime, has been developed in \cite{Mello} considering the effect of the electrostatic self-interaction on the charged particle caused by the non-trivial topology of the spacetime \cite{Mello2}. Moreover, the charged particle-dyon system has been analyzed on a conical spacetime in a nonrelativistic approach \cite{Mello1}. In these publications, the magnetic monopole and dyon were considered superposed to the respective topological defects.

According to the modern concepts of theoretical physics, topological defects may have been formed by the vacuum phase transition in the early Universe \cite{Kibble,Vilenkin}. These include domain wall, cosmic strings and monopoles. Among them, cosmic strings and monopoles seem to be the best candidate to be observed. Cosmic strings \cite{Vilenkin1} and global monopoles \cite{BV} are exotic topological objects, they do not produce local gravitational interaction, however they modify the geometry of the spacetime producing planar and solid angle deficit, respectively. In \cite{Linet}, Linet showed that a cosmic string spacetime can be produced by the vortex system, i.e., a system composed by charged scalar field and Abelian gauge field which undergoes to spontaneous symmetry breaking. Admitting that the parameter $\lambda$, associated with the scalar self-interaction potential, and electric charge $e$, both go to infinity, keeping, however, the relation $e^2=8\lambda$, the geometry produced by this system is given by the following line element below expressed in cylindrical coordinates
\begin{eqnarray}
\label{cs}
	ds^2=-dt^2+dz^2+d\rho^2+b^2\rho^2d\phi^2 \ ,
\end{eqnarray}
with $\rho\geq 0$ and $0\leq\phi\leq 2\pi$. In the expression above, $b$ is a parameter smaller than unity which depends on the linear energy density of the vortex. In the axis $\rho=0$ there exist a linear magnetic flux. 

The global monopole is a spherically symmetric topological object formed by system composed by a self-coupling scalar triplet whose original global $O(3)$ symmetry is spontaneously broken to $U(1)$. The influence of this object on the geometry of the spacetime can be evaluated by coupling the energy-momentum tensor associated with this system with the Einstein equation. Admitting the most general static spherically symmetric metric
tensor, given by the line element below,
\begin{equation}
ds^2=-B(r)dt^2+A(r)dr^2+r^2(d\theta^2+\sin^2\theta d\phi^2) \ ,
\end{equation}
Barriola and Vilenkin \cite{BV} found regular solutions for the radial functions $B(r)$ and $A(r)$, that for points far from the monopole's core read \begin{equation}
B(r)=A^{-1}(r)=1-8\pi G\eta^2-2GM/r \ ,
\end{equation}
$\eta$ being the scale energy where the symmetry is broken and $G$ the gravitational constant. The parameter $M$ is approximately the mass of the monopole. Neglecting the mass term and rescaling the time variable, we can rewrite the monopole metric tensor as
\begin{equation}
\label{gm}
ds^2=-d t^2+\frac{dr^2}{\alpha^2}+r^2(d\theta^2+\sin^2\theta d\phi^2) \ ,
\end{equation}
where the parameter $\alpha^2=1-8\pi G\eta^2$ is smaller than unity.

In $1931$, P. M. Dirac, with the objective to obtain a global dual symmetry between electric and magnetic fields in the Maxwell equations, proposed the existence of the magnetic monopole \cite{Dirac}. In his formulation the vector potential associated with the monopole is regular everywhere except on a semi-infinite line starting from the monopole. Few years later Wu and Yang \cite{WY}, proposed an elegant formalism to describe magnetic monopole free of singular line. In their formalism the vector potential is defined in two overlapping regions, $R_a$ and $R_b$, which cover the whole space. Using spherical coordinate system, with the monopole at origin they defined:
\begin{eqnarray}
\label{Ra}
R_a&:& 0 \leq \theta < \frac12\pi+\delta, \  r>0, \  0\leq \phi 
<2\pi \ , \\
R_b&:& \frac12\pi-\delta< \theta \leq \pi, \ r>0, \ 0\leq \phi < 2\pi ,\\ 
\label{Rab}
R_{ab}&:& \frac12\pi-\delta< \theta< \frac12\pi+\delta, \ r>0,  \
0\leq \phi < 2\pi \ , 
\end{eqnarray}
with $0<\delta \leq\frac12\pi$, and $R_{ab}$ being the overlapping region.

The only non-vanishing components of vector potential are
\begin{eqnarray}
\label{A}
(A_\phi)_a&=&g(1-\cos\theta) \ ,
\nonumber\\
(A_\phi)_b&=&-g(1+\cos\theta) \ ,
\end{eqnarray}
$g$ being the monopole strength. In the overlapping region the non-vanishing components are related by a gauge transformation
\begin{equation}
(A_\phi)_a=(A_\phi)_b+\frac ieS\partial_\phi S^{-1} \ ,
\end{equation}
where $S=e^{2iq\phi}$. In order to have single valued gauge transformation, we must have $q=eg=n/2$.

The well known prescription adopted to introduce electromagnetic four-vector potential, $A_\mu$, in the Dirac and Klein-Gordon equations is by the minimal coupling one. By this prescription, the four-vector differential operator $P_\mu=i\partial_\mu$ is modified as $P_\mu \longrightarrow P_\mu-eA_\mu$.  Several years ago, Dosch, Jansen and M\"uller in \cite{DJM}, pointed out that the minimal coupling is not the only way to couple a potential to the Dirac equation. There, it was suggested that a non-electromagnetic potential can be taken into account by making a modification in the mass term as $M\longrightarrow M+S({\vec{r}},t)$, being $S({\vec{r}},t)$ the named scalar potential. This new formalism has been used by Soff, M\"uller, Rafelski and Greiner, in \cite{GBW}, to analyse the Dirac equation in the presence of a Coulomb potential and a static scalar potential proportional to the inverse of the radial distance, i.e., $S\propto 1/r$. Recently the introduction of the self-interaction on a charged particle in the cosmic string spacetime \cite{Linet1,Smith}\footnote{The electrostatic self-interaction on a charged particle in the spacetime of a cosmic string is a consequence of the distortion on the electric fields caused by the planar angle deficit produced by the defect. This self-interaction is proportional to the inverse of the distance from the particle to the cosmic string.} has been used in \cite{Nail} and \cite{Spi} to analyse its  relativistic quantum motion. 

In this paper we shall analyse the relativistic quantum motion of a spin$-0$ massive charged particle in the presence of a dyon, Aharonov-Bohm magnetic flux and scalar potential, in the spacetimes produced by an idealized cosmic string and by a point-like global monopole, separately. We shall assume the dyon and the linear magnetic flux are superposed to both gravitational defects; moreover two specific configurations to the scalar potential will be considered: $i)$ the potential proportional to the inverse of the radial distance, $1/r$, and $ii)$ the potential linear to the radial distance, $r$. With these potentials we are able to obtain exact solutions to the Klein-Gordon equation, providing the energy spectra associated with bound states, and the phase shifts for scattering states. In \cite{VV}, Villalba, by using the Dirac equation, analyzed the relativistic quantum motion of a massive charged spin$-1/2$ particle in the presence of Coulomb, scalar potential and the magnetic fields of a monopole and Aharonov-Bohm flux. Considering that the scalar potential couples with the field by modifying the mass term in the Dirac equation as $m$ to $m+S(r)$, with $S(r)=-\frac{\alpha'}r$, the author found an algebraic expression to the energy spectrum. It is shown that in the specific case where the Coulomb potential is turned off, the energy spectrum assumes positive and negative values as well.

The Klein-Gordon equation for a massive charged particle, in the presence of external electromagnetic and scalar potentials presents the following form:
\begin{eqnarray}
\label{KG}
	[{\cal D}^2-\xi R-(M+S(r))^2]\Psi(x)=0\ ,
	\label{14}
\end{eqnarray}
where the differential operator above is given by
\begin{eqnarray}
\label{D}
	{\cal D}^2=\frac{1}{\sqrt{-g}}D_\mu(\sqrt{-g}\,g^{\mu\nu}D_\nu) \ ,
\end{eqnarray}
with $D_\mu=\der_\mu-ieA_\mu$. Moreover $g=det(g_{\mu\nu})$ and $S(r)$ is the scalar potential. Notice that we have introduced a non-minimal coupling, $\xi$, between the field with the scalar curvature, $R$. The electromagnetic fields that we shall consider are due to the dyon and the Aharonov-Bohm flux. In this case the four-vector potential, in spherical coordinates, reads:
\begin{eqnarray}
	A_\mu=(A_0,0,0,A_\phi) \ ,
\end{eqnarray}
being
\begin{eqnarray}
	A_\phi=\frac{\Phi_B}{2\pi}+A_\phi^{\mbox{g}}\ .
	\label{potvet}
\end{eqnarray}
$\Phi_B$ denoting the Aharonov-Bohm flux, and $A_\phi^{\mbox{g}}$ given by (\ref{A}). The Coulomb potential produced by the electric charge, $Q$, of the dyon is:
\begin{eqnarray}
	A_0(r)=\frac{Q}r \ .
\end{eqnarray}

This paper is organized as follows: In Sec. $2$ we analyse the system admitting that the spacetime is produced by a cosmic string. Because the electric and magnetic fields produced by the dyon present spherical symmetry, we shall write the metric tensor (\ref{cs}) in spherical coordinates. Considering stationary states, $\Psi({\vec{r}},t)=e^{-iEt}\Psi_E({\vec{r}})$, with $\Psi_E({\vec{r}})=R(r)Y(\theta,\phi)$, we first present the general solution to the angular function, $Y$, named {\it conical monopole harmonics}. The next steps is to calculate the radial function considering for the scalar potential two distinct expressions: $S(r)=\frac{\eta_C}r$ and $S(r)=\eta_L r$. For each case, bound and scattering states are analysed. In Sec. $3$ we develop a similar analysis considering at this time the spacetime produced by a global monopole, whose metric tensor is given by (\ref{gm}).  Finally we leave for Sec. $4$ our conclusions and most relevant remarks. 

\section{Quantum Mechanical Analysis in the Cosmic String Spacetime}

Due to the spherical symmetry associated with the vector potential (\ref{A}) and also to the scalar potentials considered, we decided to analyse the charged particle-dyon system by using spherical coordinates system. In this system the metric tensor associated with the spacetime produced by an idealized cosmic string reads
\begin{equation}
\label{cs1}
ds^2=-dt^2+dr^2+r^2d\theta^2+b^2r^2\sin^2\theta d\phi^2 \ .
\end{equation}
So, in this spacetime the differential operator (\ref{D}) becomes
\begin{eqnarray}
	{\cal D}^2&=&-(\der_t-ieA_0)^2+\frac1{r^2}\der_r(r^2\der_r)+\frac1{r^2\sin\th}\der_\th(\sin\th\der_\th)\nonumber\\
&+& \frac{(\der^2_\phi-2ieA_\phi\der_\phi-e^2A_\phi^2)}{b^2r^2\sin^2\th} \ .
	\label{D1}
\end{eqnarray}

According to the previous analysis by Wu and Yang \cite{WY1}, solutions to the Klein-Gordon equation in presence of the Wu-Yang magnetic monopole field will not be ordinary functions but, instead, {\it sections}, i.e., the solutions assume values $\Psi_a$ and $\Psi_b$ in $R_a$ and $R_b$, and satisfy the gauge transformation
\begin{equation}
\Psi_a=S\Psi_b  \ 
\end{equation}
in the overlapping region $R_{ab}$. 

In order to analyse the quantum motion of the particle, let us admit that the wave-function, solution of (\ref{KG}), has the general form
\begin{eqnarray}
\label{P}
	\Psi(x)=e^{-iEt}R(r)Y(\th,\phi) \ ,
\end{eqnarray}
$E$ being the self-energy of the particle.

Substituting (\ref{P}) and (\ref{D1}) into (\ref{KG}), and remembering that the curvature scalar, $R$, vanishes for this spacetime, we obtain two differential equations: $i)$ the equation associated with the angular variables,
\begin{eqnarray}
	\left[\frac1{\sin\th}\der_\th(\sin\th\der_\th)+\frac{(\der^2_\phi-2ieA_\phi\der_\phi-e^2A_\phi^2)}{b^2\sin^2\th}\right]Y(\th,\phi)=- \lam Y(\th,\phi)
	\label{eqangular}
\end{eqnarray}
and, $ii)$ the equation for the radial function,
	\begin{eqnarray}
		\frac1{r^2}\frac{d}{dr}\left(r^2\frac{dR}{dr}\right)-\left[\frac{\lam}{r^2}+(M+S(r))^2 -(E+eA_0)^2\right]R=0\ .
		\label{Rad}
	\end{eqnarray}
In both equations $\lam$ is a constant factor introduced to separate the original differential equation. 

\subsection{Conical Monopole Harmonics}	
In order to obtain solution for (\ref{eqangular}), we have to consider this equation in the regions $R_a$ and $R_b$ separately. Following the procedure adopted in \cite{WY1} we can assume that
\begin{equation}
Y(\theta,\phi)=\Theta(\theta)e^{i(m\pm q)\phi}\ ,
\label{Y}
\end{equation}
where the positive (negative) sign refers to region $R_a$ ($R_b$). $m$ is the magnetic quantum number and $q=eg$. 

The solution of (\ref{eqangular}) in absence of the Aharonov-Bohm magnetic field, has been obtained in \cite{Mello1}. This solution is a generalization of the monopole harmonics defined by Wu and Yang in \cite{WY1} to conical spacetime, that we named {\it conical monopole harmonics}. So below we  briefly reproduce some steps adopted to provide the solution in the present case, i.e., in the presence of a magnetic flux along the cosmic string. Substituting (\ref{Y}) into (\ref{eqangular}), and introducing a new variable $x=\cos\theta$, we obtain for both regions $R_a$ and $R_b$ the same differential equation
\begin{eqnarray}
(1-x^2)\frac{d^2\Th(x)}{dx^2}-2x\frac{d\Th(x)}{dx}-\frac{(m_{\Phi,b}+q_b x)^2}{(1-x^2)}\Th(x) =-\lam\Th(x)\ ,
\label{Te}
\end{eqnarray}
where\footnote{In our development we define the ratio of the flux $\Phi_B$ to the quantum flux, $\phi_0=2\pi/e$, by $\frac{\Phi_B}{\Phi_0}= {\bar{N}}+\epsilon$, being ${\bar{N}}$ an integer number and $0<\epsilon<1$.}
\begin{eqnarray}
	q_b=\frac{q}{b} \ , \  \mbox{and} \ \ \ m_{\Phi,b}=\frac{m-({\bar{N}}+\epsilon)}b \ .
\end{eqnarray}
The solution of the above differential equation can be expressed in terms of the Jacobi polynomials \cite{Grad} by 
\begin{eqnarray}
	\Th(x)=(1-x)^{\sigma/2}(1+x)^{\beta/2}P_n^{\,\sigma,\beta}(x)\ , 
\end{eqnarray}
$P_n^{\,\sigma,\beta}(x)$ being the Jacobi polynomials of degree $n$ expressed by
\begin{eqnarray}
	P_n^{\,\sigma,\beta}(x)=\frac{(-1)^n}{2^n\,n!}(1-x)^{-\sigma}(1+x)^{-\beta} \frac{d^{\,n}}{dx^{\,n}}\left[(1-x)^{\sigma+n}(1+x)^{\beta+n}\right],
\end{eqnarray}
with
\begin{eqnarray}
	\sigma=m_{\Phi,b}+q_b , \ \ \ \beta=m_{\Phi,b}-q_b ,\ \ \ n=l_{\Phi,b}-m_{\phi,b}\ .
\end{eqnarray}

The eigenvalues of (\ref{Te}) are given by 
\begin{eqnarray}
\lam\equiv\lam_{\phi,b}=l_{\phi,b}(l_{\phi,b}+1)-q_b^2\ ,
\end{eqnarray}
being 
\begin{eqnarray}
	l_{\Phi,b}=l-({\bar{N}}+\epsilon)+\left(\frac1b-1\right)\left|m-({\bar{N}}+\epsilon)\right|.
\end{eqnarray}

Because $l_{\phi,b}(l_{\phi,b}+1)\geq q_b^2$ we must have $l_{\phi,b}\geq \frac{q}b$. In this way the possibles values to $l$ are $l=|q|+{\bar{N}}+1,\ |q|+{\bar{N}}+2 \ ...$.

Finally the normalized solution of (\ref{eqangular}), in the region $R_a$, is
\begin{eqnarray}
\label{MH}
	Y^{q_b}_{l_b,m_b}(\th,\phi)=N_{q_b,l_b}(1-x)^{\sigma/2}(1+x)^{\beta/2}P_n^{\,\sigma,\beta}(x)e^{i(m+q)\phi},
\end{eqnarray}
with the normalization constant 
\begin{eqnarray}
N_{q_b,l_b}=\frac1{\sqrt{2\pi b}}\left[\frac{(2n+\sigma+\beta+1)n!\Ga(n+\sigma+\beta+1)} {2^{\sigma+\beta+1}\Ga(n+\sigma+1)\Ga(n+\beta+1)}\right]^{1/2}\ .
\end{eqnarray}
So the conical monopole harmonics in the presence of the Aharonov-Bohm magnetic flux obey the eigen-values equation
\begin{eqnarray}
\label{YY}
	{\vec L}^{\,2}_{q_b}Y^{q_b}_{l_{\Phi,b},\, m_{\Phi,b}}(\th,\phi)=l_{\Phi,b}(l_{\Phi,b}+1)Y^{q_b}_{l_{\Phi,b},\, m_{\Phi,b}}(\th,\phi)\ ,
	\end{eqnarray}
being
\begin{eqnarray}
\label{LY}
	{\vec L}^{\,2}_{q_b}=-\frac1{\sin\th}\der_\th(\sin\th\der_\th)- \frac1{b^2\sin^2\th}\left(\der_\phi-ieA_\phi\right)^2+q_b^2\ .
	\label{angular}
\end{eqnarray}
	
\subsection{Radial Equation}
Now let us return to the analysis of the radial equation (\ref{Rad}), where now we have the value of the parameter $\lam\equiv\lam_{\phi,b}=l_{\phi,b}(l_{\phi,b}+1)-q_b^2$. In what follows, this analysis will be developed in the two next subsections considering two different configurations to the scalar potential. At this point we want to call attention that with these configurations, exact solutions to the radial differential equations can be obtained; moreover, these configurations may be considered to analyze the relativistic motion of a particle in the presence of a self-interaction potential due to the geometry of the spacetime, and in the presence of an isotropic harmonic oscillator. 

\subsubsection{Scalar Potential $S(r)=\frac{\eta_C}r$}	
Considering that the scalar potential is Coulomb-type,
\begin{eqnarray}
 	S(r)=\frac{\eta_C}r \ ,
\end{eqnarray}
the radial equation presents the following form:
\begin{eqnarray}
\frac1{r^2}\frac{d}{dr}\left(r^2\frac{dR}{dr}\right)+\left[\frac{e^2Q^2-\eta_C^2-\lam_{\phi,b}}{r^2}+2\frac{eQE-M\eta_C}r+E^2-M^2\right]R=0 \ .
\end{eqnarray}
For this case the system may have bound and scattering states. 

In order to have bound states, it is necessary that $E<M$, the product $eQ$ be a positive quantity and the parameter $\eta_C< eQ$\ \footnote{If we assume that the product $eQ$ is a negative quantity, the parameter $\eta_C$ should be negative and its modulus greater than $|eQ|\frac{E}{M}$. Because in this case the value of $\eta_C$ will depend on the ratio of energy by the mass of the particle, we shall not consider this situation.}. Admitting this situation, the solution for the radial equation is:
\begin{eqnarray}
	R(r)=e^{-kr}(kr)^{s_b-1}{_1}F_1\left(s_b+\frac\nu2,2s_b; 2kr\right) ,
\end{eqnarray}
with
\begin{eqnarray}
\label{def}
	\mu_{\phi,b}=\lam_{\phi,b}+\eta_C^2-e^2Q^2 , \ \ \ \nu=2\frac{M\eta_C-eQE}k , \ \ \ k^2=M^2-E^2
\end{eqnarray}
and
\begin{eqnarray}
	s_b=\frac12+\frac{\sqrt{1+4\mu_{\phi,b}}}2 \ .
\end{eqnarray}

Because the divergent behavior of the hypergeometric function ${_1}F_1$ for large values of its argument, bound states solution can only be obtained by imposing that this function becomes a polynomial of degree $N$. In this case the radial solution above goes to zero at infinity. This condition is obtained by imposing 
\begin{eqnarray}
\label{s}
	s_b+\frac\nu2=-N\, ,\ \ \ \ N=0,1,2,3,...\ .
\end{eqnarray}
Combining conveniently the equations given in (\ref{def}) and (\ref{s}), we obtain\footnote{Similar expressions have been obtained by Soff {\it at al} in \cite{GBW} and by Villalba in \cite{VV}, by using the Dirac equation to analyse the quantum motion of a charged particle in the presence of Coulomb and scalar potential.}
\begin{eqnarray}
\label{EM}
	(s_b+N)\sqrt{M^2-E^2}=eQE-\eta_C M \ .
\end{eqnarray}
This equation provides the quantization condition on the energy spectrum of the particle:
\begin{eqnarray}	E_{N,\phi,b}=M\left[\frac{eQ\eta_C\pm\left(N+s_b\right)\sqrt{e^2Q^2-\eta_C^2+\left(N+s_b\right)^2}}{e^2Q^2+\left(N+s_b\right)^2}\right] \ .
\end{eqnarray}

Two particular situations deserve to be mentioned: $i)$ $\eta_C=0$, which corresponds the presence of bound states due to the Coulomb potential of the dyon only. In this case the energy spectrum becomes
\begin{eqnarray}
E=M\left[1+\frac{e^2Q^2}{\left(N+s_b\right)^2}\right]^{-1/2} \ .
\end{eqnarray}
The negative sign has been discarded because it is incompatible with the condition (\ref{EM}).

The second possibility is given for $ii)$ $Q=0$, which corresponds the absence of the Coulomb potential of the dyon. In this case bound states are due to the presence of the scalar potential and the energy spectrum becomes
\begin{eqnarray}
E=\pm M\left[1-\frac{\eta_C^2}{\left(N+s_b\right)^2}\right]^{1/2}\, .
\end{eqnarray}
For this case both sign for the energy are allowed.

After this analysis, let us study scattering states. Scattering states can be obtained by imposing the condition $E>M$ in the solution of the radial differential equation. In this case we get:
\begin{eqnarray}
	R(r)=e^{i\kappa r}(\kappa r)^{s_b-1}{_1}F_1\left(s_b+\frac{i\nu}2,2s_b; -2i\kappa r\right) ,
\end{eqnarray}
where
\begin{eqnarray}
\kappa^2=E^2-M^2\ , \ \ \ \ \ \nu=2\frac{M\eta_C-eQE}\kappa\, .
\end{eqnarray}

By the asymptotic behavior of $R(r)$ for $\kappa r>>1$, it is possible to obtain the expression to the phase shift, $\delta_l$, the most important parameter to calculate the scattering amplitude. The behavior of the radial function for large value of its argument is
\begin{eqnarray}
\label{ra}
	R(r)\approx\frac1{\kappa r}\cos\left[\kappa r-\frac\nu2\ln(2\kappa r)+\gamma_l- \frac\pi2 s_b\right] \ ,
\end{eqnarray}
where
\begin{eqnarray}
\gamma_l=\arg\Ga\left(s_b+\frac{i\nu}2\right)\ .
\end{eqnarray}

So, we can see that the phase shift presents two distinct contributions\footnote{The logarithmic term in (\ref{ra}) is consequence of the long range interaction of the Coulomb and scalar potentials. Because it does not depend on the partial wave analysis and varies slowly with the distance, it cannot be considered as a contribution for $\delta_l$ \cite{Schiff,Merz}.}:\\
$i)$ One due to the Coulomb and scalar potential
\begin{eqnarray}
	\delta^{(1)}_l=\gamma_l=\arg\Ga\left(s_b+\frac{i\nu}2\right)\ ,
\end{eqnarray}
which disappears for $Q=\eta_C=0$, and\\
$ii)$ the other from the modification of the effective angular quantum number, $\lam_{\phi,b}$, caused by the geometry of the spacetime, by the interaction with the magnetic and electric fields of the dyon, by the magnetic field of the Aharonov-Bohm flux, and finally there exist a contribution due to the scalar potential:
\begin{eqnarray}
	\delta^{(2)}_l=\frac\pi2 (l+1-s_b)\ .
\end{eqnarray}
So the complete phase shift is give by the sum of both phases
\begin{eqnarray}
	\delta_l=\delta^{(1)}_l+\delta^{(2)}_l\ .
\end{eqnarray}

\subsubsection{Scalar Potential $S(r)=\eta_L r$}	

In this case the radial differential equation (\ref{Rad}) becomes
\begin{eqnarray}
\frac1{r^2}\frac{d}{dr}\left(r^2\frac{dR}{dr}\right)+\left[\frac{e^2Q^2-\lam_{\phi,b}}{r^2}+2\frac{eQE}r-2M\eta_L r-\eta_L^2r^2+E^2-M^2\right]R=0 \ .
\end{eqnarray}
Th above equation is similar to the non-relativistic one which governs the quantum motion of a charged particle in the presence of a dyon and an extra isotropic harmonic potential\footnote{In \cite{Mello1} the non-relativistic quantum motion of a charged particle in the presence of a dyon, and an isotropic harmonic potential is analyzed.}. In this way the parameter $\eta_L$ can be identified as $M\omega$, being $\omega$ the classical frequency of the oscillator. Introducing a new function $R(r)=u(r)/r$ and defining a new variable $z=\sqrt{\eta_L}r$, the above equation reads
\begin{eqnarray}
\frac{d^2u}{dz^2}+\left[\frac{e^2Q^2-\lam_{\phi,b}}{z^2}+\frac{2eQE}{\eta_L^{1/2}z}- \frac{2M}{\eta_L^{1/2}}z-z^2+\frac{E^2-M^2}{\eta_L}\right]u=0\ .
	\label{RR}
\end{eqnarray}

In order to investigate the solutions of the above differential equation, it is useful to analyse its behavior in the regions $z\rightarrow0$ and $z\rightarrow\infty$. Doing this we verify that $u(z)$ presents the general form below:
\begin{eqnarray}
	u(z)=z^{\beta/2}e^{-z(z+c)/2}F(z)\, ,
	\label{uu}
\end{eqnarray}
with
\begin{eqnarray}
	c=\frac{2M}{\eta_L^{1/2}} \ , \ \ \ \beta=1+\sqrt{1+4(\lam_{\phi,b}-e^2Q^2)}
\end{eqnarray}
and $F(z)$ satisfying the following differential equation
\begin{eqnarray}
\label{F}
	zF''(z)+(\beta-cz-2z^2)F'(z)-[a_1+a_2z]F(z)=0\, ,
\end{eqnarray}
with
\begin{eqnarray}
	a_1=\frac{c\beta}2-\frac{2eQE}{\eta_L^{1/2}}\ \ \ \ \mbox{and}\ \ \ \ a_2=\beta-\frac{E^2}{\eta_L}\ .
\end{eqnarray}

Exact solutions of (\ref{F}) can be obtained admitting a series expansion for the unknown function: 
\begin{eqnarray}
	F(z)=\sum_{k=0}^\infty d_kz^k \ .
\end{eqnarray}
Substituting the series above into (\ref{F}) we obtain the following recurrence relations:
\begin{eqnarray}
	d_1=\frac{a_1}\beta d_0 \ \ \ 
\end{eqnarray}
and 
\begin{eqnarray}
  d_{k+2}=\frac{(k+1)c+a_1}{(k+2)(k+1+\beta)}d_{k+1}+\frac{2k+a_2}{(k+2)(k+1+\beta)}d_k\ .
  \label{recor}
\end{eqnarray}
The  analysis of a differential equation similar to (\ref{F}) was develop by Ver\c cin \cite{Ver} and Ver\c cin {\it at al} \cite{Ver1}, investigating the quantum planar motion of two anyons with Coulomb interaction between them and in the presence of an external uniform magnetic field. In the first paper the author found exactly the energy spectrum, by imposing specific conditions on the wave-function. In the second,  the energy spectrum was evaluated numerically. Moreover a similar equation has also obtained in \cite{Fur}, analyzing the energy spectrum of an electron or hole in the presence of a magnetic field in the framework of the theory of defect of Katanaev and Volovich \cite{KV}, and in \cite{Bezerra}, analyzing the nonrelativistic problem of two charged particle on a cone in the presence of a static uniform magnetic field. 

Now let us return to the series solution to the function $F(z)$. Admitting $d_0=1$ and using (\ref{recor}), we obtain
\begin{eqnarray}
	d_2 &=&\frac{a_1(c+a_1)+a_2\beta}{2\beta(\beta+1)}\nonumber\\
	d_3&=&\frac{(2c+a_1)(c+a_1)a_1}{3!\beta(\beta+1)(\beta+2)}+ \frac{(2c+a_1)a_2\beta}{3!\beta(\beta+1)(\beta+2)}+ \frac{a_1(2+a_2)}{3\beta(\beta+2)} \ .
\end{eqnarray}

Special kind of exact solutions, which represent bound states, can be obtained looking for polynomials expressions to $F(z)$. Being $n$ the order of such polynomial, two different conditions must be satisfied simultaneously: $a_2=-2n$ and $d_{n+1}=0$. In this way we have $d_{n+2}=d_{n+3}=...=0$. The condition $a_2=-2n$ provides the energy spectrum for the system, $E_n^2=(\beta+2n)\eta_L$, and the condition $d_{n+1}=0$ an algebraic equation satisfied by specific values of the parameter $\eta_L$.

In what follows we shall consider the simplest case where the function $F$ is a polynomial of first order. In this case we have:\\
$i)$ The condition $a_2=-2$ implies
\begin{eqnarray}
	F(z)=1+\frac{a_1}{\beta} z\ .
\end{eqnarray}
The energies associated with this states are
\begin{eqnarray}
	E_1=\pm\sqrt{(\beta+2)\eta_L}\ .
\end{eqnarray}
Both sign for the energy are allowed for this system.\\
$ii)$ The condition $d_2=0$ restrict the possible values that the parameter $\eta_L$ can have. For this case these values are
\begin{eqnarray}	
\label{eta1}
\eta_L^{\pm}&=&\frac{(\beta+2)M^2}2\left[\frac{2e^2Q^2[(\beta+1)^2+1]+\beta^2}{[2e^2Q^2(\beta+2)-\beta]^2}\right.\nonumber\\	&\pm&\left.\frac{2eQ(\beta+1)\sqrt{2(2e^2Q^2+\beta^2)}}{[2e^2Q^2(\beta+2)-\beta]^2}\right]\ .
\end{eqnarray}

From these results we can see that: $i)$ The two distinct values to $\eta_L$ are positive, consequently, the energy spectrum is twice bigger than in the case of just one solution. $ii)$ Bound states solutions exist for both sign of the product $eQ$; moreover the associated energy depends on the parameter $b$ which codify the presence of the cosmic string, and on the ratio of the Aharonov flux by the quantum flux, $\frac{\Phi}{2\pi/e}$. $iii)$ Finally we should say that there exist infinite series solution for $F(z)$ with physically acceptable behavior \cite{Ver1}, however, for these cases the energy spectrum can only be obtained numerically. 

\section{Quantum Mechanical Analysis in the Global Monopole Spacetime}
The metric tensor of the spacetime produced by a point-like global monopole is given by
\begin{equation}
ds^2=-d t^2+\frac{dr^2}{\alpha^2}+r^2(d\theta^2+\sin^2\theta d\phi^2) \ .
\end{equation}
In this spacetime the differential operator (\ref{D}) reads
\begin{eqnarray}
	{\cal D}^2&=&-(\der_t-ieA_0)^2+\frac{\al^2}{r^2}\der_r(r^2\der_r)-\nonumber \\
	&-&\frac1{r^2}\left[-\frac1{\sin\th}\der_\th(\sin\th\der_\th)-\frac1{\sin^2\th}\der^2_\phi+\frac{2ieA_\phi}{\sin^2\th}\der_\phi+ \frac{e^2A_\phi^2}{\sin^2\th}\right]\nonumber\\
	&=&-(\der_t-ieA_0)^2+\frac{\al^2}{r^2}\der_r(r^2\der_r)-\frac{{\vec L}^2_q}{r^2}\ ,
	\label{015}
\end{eqnarray}
${\vec L}^2_q$ being the operator defined in (\ref{LY}) taking $b=1$, whose eigen-functions are the monopole harmonics in the presence of the Aharonov-Bohm flux, given in (\ref{MH}). We shall admit for the wave-function the following form:
\begin{eqnarray}
	\Psi(x)=e^{-iEt}R(r)Y^q_{l_\phi,m_\phi}(\th,\phi) \ .
	\label{016}
\end{eqnarray}

Substituting (\ref{015}) and (\ref{016}) into (\ref{14}), and considering for the scalar curvature $R=2(1-\al^2)/r^2$, we obtain the following radial differential equation
\begin{eqnarray}
	\frac{\al^2}{r^2}\frac{d}{dr}\left(r^2\frac{dR}{dr}\right)-\left[\frac{\lam_{\phi}+2\xi(1-\al^2)}{r^2}+(M+S(r))^2-(E+eA_0)^2\right]R=0\ ,
	\label{0017}
\end{eqnarray}
with $\lam_\phi=l_\phi(l_\phi+1)-q^2$.

As in the previous section we shall consider, separately, two distinct configurations for the scalar potential.

\subsection{Scalar Potential $S(r)=\frac{\eta_C}r$}	
Considering for the scalar potential the Coulomb-type one\footnote{It has been shown in \cite{Mello2} that a self-interaction on a charged particle placed in the spacetime of a point-like global monopole is given by $U=\frac{K(\alpha)}{r}$, $K(\alpha)$ being a positive constant for $\alpha<1$ and $r$ the radial distance to the monopole.}
\begin{eqnarray}
 	S(r)=\frac{\eta_C}r \ 
\end{eqnarray}
the radial equation becomes:
\begin{eqnarray}
\frac1{r^2}\frac{d}{dr}\left(r^2\frac{dR}{dr}\right)&+&\left[\frac{e^2Q^2-\eta^2_C-2\xi(1-\al^2)-\lam_\phi}{\al^2r^2}+\right.\nonumber \\
&+&\left.2\frac{eQE-M\eta_C}{\al^2r}+\frac{E^2-M^2}{\al^2}\right]R=0 \ .
\end{eqnarray}
This system also presents bound and scattering states.

Bound states take place in the system if we assume $E<M$ and $\eta_C<eQ$, being $eQ$ a positive quantity. Admitting this situation the solution to the radial differential equation above is
\begin{eqnarray}
\label{Rrr}
	R(r)=e^{-kr}(kr)^{s_\alpha-1}{_1}F_1\left(s_\alpha+\frac\nu2,2s_\alpha; 2kr\right) ,
\end{eqnarray}
with
\begin{eqnarray}
	\mu_{\phi,\al}=\frac{\lam_\phi+\eta_C^2+2\xi(1-\al^2)-e^2Q^2}{\al^2} , \ \ \ \nu=2\frac{M\eta_C-eQE}{\al^2k} , \ \ \ k^2=\frac{M^2-E^2}{\al^2}
	\label{0018}
\end{eqnarray}
and
\begin{eqnarray}
	s_\alpha=\frac12+\frac{\sqrt{1+4\mu_{\phi,\alpha}}}{2} \ .
\end{eqnarray}

As in the previous analysis, to obtain bound states it is necessary that the hypergeometric function becomes a polynomial on degree $N$. This condition is attained by imposing
\begin{eqnarray}
	s_\alpha+\frac\nu2=-N \ .
	\label{0019}
\end{eqnarray}
Combining conveniently the equations given in (\ref{0018}) and (\ref{0019}), we obtain
\begin{eqnarray}
	\al^2(s_\alpha+N)\sqrt{M^2-E^2}=eQE-\eta_C M \ ,  
\end{eqnarray}
which provides the quantization condition on the energy spectrum of the system:
\begin{eqnarray}	E=M\left[\frac{eQ\eta_C\pm\al^2\left(N+s_\alpha\right)\sqrt{e^2Q^2-\eta^2_C+\al^4\left(N+s_\alpha\right)^2}}{e^2Q^2+\al^4\left(N+s_\alpha\right)^2}\right] \ .
\end{eqnarray}

Two distinct situations can be analysed: $i)$ Considering $\eta_C=0$. In this case the Coulomb attractive potential of the dyon provides an energy spectrum 
\begin{eqnarray}
\label{EM1}
E=M\left[1+\frac{e^2Q^2}{\al^4\left(N+s_\alpha\right)^2}\right]^{-1/2}\, .
\end{eqnarray}
The second possibility is given for $ii)$ $Q=0$. In this case the scalar potential can bind the particle if $\eta_C<0$, and the energy spectrum is given by\footnote{The same considerations about the sign of the energy spectrum mentioned in the last section take place in this analysis.}
\begin{eqnarray}
\label{EM2}
E=\pm M\left[1-\frac{\eta^2_C}{\al^4\left(N+s_\alpha\right)^2}\right]^{1/2}\, .
\end{eqnarray}

Scattering states can be obtained admitting that $E>M$ in the radial differential equation. The respective solution is achieved by taking $k=-i\kappa$ in (\ref{Rrr}). The solution is
\begin{eqnarray}
	R(r)=e^{i\kappa r}(\kappa r)^{s_\alpha-1}{_1}F_1\left(s_\alpha+\frac{i\nu}2,2s_\alpha; -2i\kappa r\right)\, .
\end{eqnarray}
The asymptotic behavior of this function for $\kappa r>>1$ is given by
\begin{eqnarray}
	R(r)\approx\frac1{\kappa r}\cos\left[\kappa r-\frac\nu2\ln(2\kappa r)+\gamma_l-\frac\pi2s_\alpha\right]
\end{eqnarray}
being
\begin{eqnarray}
	\gamma_l=\arg\Ga\left(s_\alpha+\frac{i\nu}2\right)\ .
\end{eqnarray}

As in the previous section, this asymptotic behavior indicates that the phase shift, $\delta_l$, presents two distinct contributions:\\
$i)$ One due to the Coulomb and scalar potential
\begin{eqnarray}
	\delta^{(1)}_l=\gamma_l=\arg\Ga\left(s_\alpha+\frac{i\nu}2\right)\ .
\end{eqnarray}
and\\
$ii)$ the other due to the modification of the angular quantum number caused by the geometry of the spacetime, by the electromagnetic interaction and the presence of the scalar potential
\begin{eqnarray}
	\delta^{(2)}_l=\frac\pi2 (l+1-s_\alpha)\ .
\end{eqnarray}
Finally the complete phase shift reads
\begin{eqnarray}
	\delta_l=\delta^{(1)}_l+\delta^{(2)}_l \ .
\end{eqnarray}

\subsection{Scalar Potential $S(r)=\eta_L r$}
For this case the radial differential equation becomes
\begin{eqnarray}
\frac1{r^2}\frac{d}{dr}\left(r^2\frac{dR}{dr}\right)-\left[\frac{\mu_{\phi,\al}}{r^2}-\frac{2eQE}{\al^2r}+\frac{2M\eta_\al}{\al}r+\eta_\al^2r^2- \frac{E^2-M^2}{\al^2}\right]R=0 \ ,
\end{eqnarray}
with 
\begin{eqnarray}
	\mu_{\phi,\al}=\frac{\lam_\phi+2\xi(1-\al^2)-e^2Q^2}{\al^2}\ ,\ \ \ \ \ \ \eta_\al=\frac{\eta_L}\al\ .
\end{eqnarray}
Introducing a new function $R(r)=u(r)/r$ and defining a dimensionless variable $z=\sqrt{\eta_\al}r$, we have
\begin{eqnarray}
\label{Dug}
\frac{d^2u}{dz^2}-\left[\frac{\mu_{\phi,\al}}{z^2}-\frac{2eQE}{\al^2\eta_\al^{1/2}z}+ \frac{2M}{\al\eta_\al^{1/2}}z+z^2-\frac{E^2-M^2}{\al^2\eta_\al}\right]u=0 \ .
\end{eqnarray}

In order to obtain the energy spectrum to the above differential equation it is necessary to analyse its asymptotic behavior for $z\longrightarrow 0$ and $z\longrightarrow\infty$. After that we can express the function $u(z)$ in terms of the unknown function $F(z)$ as follows:
\begin{eqnarray}
\label{ug}
	u(z)=z^{\beta/2}e^{-z(z+c)/2}F(z)\, ,
\end{eqnarray}
where
\begin{eqnarray}
	c=\frac{2M}{\al\eta_\al^{1/2}}\, , \ \ \ \ \ \mbox{and}\ \ \ \beta=1+\sqrt{1+4\mu_{\phi,\al}} \ .
\end{eqnarray}
Substituting (\ref{ug}) into (\ref{Dug}) we obtain a differential equation for the unknown function $F(z)$ as shown below:
\begin{eqnarray}
\label{86}
	zF''(z)+(\beta-cz-2z^2)F'(z)-[a_1+a_2z]F(z)=0\, ,
\end{eqnarray}
with
\begin{eqnarray}
	a_1=\frac{c\beta}2-\frac{2eQE}{\al^2\eta_\al^{1/2}}\ \ \ \ \mbox{and}\ \ \ \ a_2=\beta-\frac{E^2}{\al^2\eta_\al} \  .
\end{eqnarray}

Using the Frobenius' method to solve (\ref{86}),
\begin{eqnarray}
	F(z)=\sum_{k=0}^\infty d_kz^k \ ,
\end{eqnarray}
we find the recurrence relations below
\begin{eqnarray}
	d_1=\frac{a_1}\beta d_0 \ \ \ 
\end{eqnarray}
and
\begin{eqnarray}
  d_{k+2}=\frac{(k+1)c+a_1}{(k+2)(k+1+\beta)}d_{k+1}+\frac{2k+a_2}{(k+2)(k+1+\beta)}d_k \ .
\end{eqnarray}

Imposing that the function $F(z)$ be a polynomial of degree $n$, two conditions must be satisfied: $a_2=-2n$ and $d_{n+1}=0$. Here, we shall apply this procedure to obtain the wave-function and respective energy to the simplest case where $F(z)$ is first order degree polynomial. In this case we get:
\begin{eqnarray}
\label{FF}
	F(z)=1+\frac{a_1}\beta z\ .
\end{eqnarray}

From the first condition, $a_2=-2$, we obtain
\begin{eqnarray}
	E_1=\pm\sqrt{\al(\beta+2)\eta_L}
\end{eqnarray}
and from the second condition, $d_2=0$, we obtain
\begin{eqnarray}
\label{eta2}	
\eta^{\pm}_L&=&\alpha\frac{(\beta+2)M^2}2\left[\frac{2e^2Q^2[(\beta+1)^2+1]+\al^2\beta^2}{[2e^2Q^2(\beta+2)-\al^2\beta]^2}\right.\nonumber\\	&\pm&\left.\frac{2eQ(\beta+1)\sqrt{2(2e^2Q^2+\al^2\beta^2)}}{[2e^2Q^2(\beta+2)-\al^2\beta]^2}\right]\, .
\end{eqnarray}
As we can see two distinct values to the parameter $\eta_L$ are possible. Each one provides different energy and wavefunction. Moreover, the energy also depends on the parameter $\alpha$ which codify the presence of the monopole, on the ratio $\frac{\Phi}{2\pi/e}$ and the curvature coupling $\xi$. As in the previous analysis, bound states exist for both sign of the product $eQ$.

\section{Concluding Remarks}
In this paper we have analysed the relativistic quantum motion of a spin$-0$ charged particle in the presence of a dyon, Aharonov-Bohm magnetic field and specific scalar potentials, $S(r)$, in the spacetime of a cosmic string and global monopole. For both spacetimes, the dyon and scalar potential are superposed to the topological gravitational defects. Two specific radial scalar potential were considered: $i)$ potential proportional to the inverse of the distance, $S(r)=\eta_C/r$, and $ii)$ potential proportional to the distance, $S(r)=\eta_L r$. For both cases we were able to find exact solutions of the Klein-Gordon equation. In the specific case of the cosmic string spacetime, we have found the conical monopole harmonics in the presence of the Aharonov-Bohm magnetic flux. For the global monopole spacetime, this solutions is promptly obtained by taking the factor $b$, associated with the string, equal to unity. Finally for both spacetimes the radial functions were also obtained. 

As to the first scalar potential, we presented bound and scattering states. By the results obtained we could see that the energy spectra associated with bound states depend on the parameter which codify the presence of the defects, on magnetic and electric charge of the dyon, the ratio of Aharonov-Bohm magnetic flux by the quantum flux, $\frac{\Phi}{2\pi/e}$, and finally on the scalar potential strength. The magnetic interactions modify the angular quantum number, $\lambda$, and the electric and scalar interactions act for binding the particle. Analyzing the asymptotic behavior of the scattering wavefunction we could present the phase shifts, $\delta_l$, and shown how they depend on the above mentioned parameter as well. In fact the phase shift presents two distinct contributions: one due to Coulomb and scalar potential, and the other, due to the modification on the effective angular quantum number. 

As to the second scalar potential, we observed that it produces radial differential equations similar to the ones obtained in a non-relativistic formalism for a charged particle in the presence of an isotropic harmonic oscillator. For this potential there are no scattering states. All the states are bound, even for repulsive Coulomb potential. In fact for $r\to\infty$, centrifugal and Coulomb potential tend to disappear while the interaction caused by the scalar potential grows without limit. Admitting a series expansion to the radial function it is possible to obtain exact solutions to the bound states and their respective self-energy. As an application of this method, we found the wavefunction and respective energy for the  ground state. In fact two different energy states were obtained for different values to the parameter $\eta_L$ given in (\ref{eta1}) and (\ref{eta2}). 

Our main objectives to include extra scalar potentials in the charged particle-dyon system were to investigate the influence on the relativistic quantum motion due to: $i)$ the presence of the non-electromagnetic self-interaction on the particle induced by the distortion on the electric fields caused by the non-trivial topology of the spacetime, and $ii)$ the presence of an extra isotropic harmonic potential acting on the particle. The latter was introduced to provide bound states between the charged particle end the dyon, even for repulsive Coulomb interaction.

At this point we would like to finish this paper by saying that monopoles or dyons have not yet been directly detected in laboratories. Experiments in great accelerators are carrying out to a direct search of them \cite{KM}. Besides, indirect search has been proposed as well \cite{Ruj,MA}. Also the possibility of bind monopole to an atomic nucleon or to a molecule, has been analysed many years ago in \cite{Th}. So the presence of these objects on the cosmos, accelerators or in composition with many atoms are not ruled out. The present paper has the objective to provide some contributions to the investigation of the relativistic quantum motion of charged particle in the presence of these objects admitting their existence in the nature. In this way, the analysis of this system in a general spacetime could be useful to understand their dynamical and consequences on the Particle Physics.

\section*{Acknowledgment} 
The authors thank Conselho Nacional de Desenvolvimento Cient\'\i fico e Tecnol\'ogico (CNPq.) and FAPESQ-PB/CNPq. (PRONEX) for partial financial support.


\end{document}